\documentclass[aps,prl,reprint,superscriptaddress,nofootinbib]{revtex4-2}
\usepackage[T1]{fontenc}
\usepackage{amsmath,amssymb,graphicx,bm,microtype}
\usepackage[colorlinks=true,citecolor=blue,urlcolor=blue]{hyperref}
\newcommand{\Li}{\mathcal L}
\newcommand{\D}{\mathcal D}
\newcommand{\Tr}{\operatorname{Tr}}

\newcommand{\EN}{E_{\mathcal N}}

\begin{document}

\title{Emergent quantum algebras in dissipative time crystals}

\author{Joseph Tindall}
\affiliation{Center for Computational Quantum Physics, The Flatiron Institute, 162 5th Avenue, New York, New York 10010, USA}
\date{\today}

\begin{abstract}
Strong dynamical symmetries provide an algebraic description of persistent oscillations in open quantum systems governed by the Lindblad equation. For finite-dimensional systems with a full-rank stationary state, such symmetries are necessary and sufficient for non-decaying oscillatory modes to appear in the Liouvillian's spectrum. These modes can be directly constructed by acting on stationary states with the identified symmetry operators. Here we establish an asymptotic counterpart of this algebra for collective spins. Taking a single spin $S$ in a magnetic field, subject to dissipation through linear jump operators, we identify the exact conditions on the field and dissipative couplings under which non-decaying oscillatory modes emerge as $S \rightarrow \infty$. We then write down a simple, closed form for the corresponding `symmetry' operator which can be applied to the steady state to construct these modes, which are eigenvectors of the Liouvillian, with imaginary eigenvalues, up to $O(1/S)$ corrections. A second operator, an emergent conserved quantity, completes the algebra, and its slow relaxation determines how quantum noise selects the stationary state. The construction can be directly extended to a pair of interacting spins sharing a common reservoir, with the resulting quantum algebra making the entanglement during a quench analytically accessible.
\end{abstract}

\maketitle

\begingroup
\let\thefootnote\relax
\footnotetext{This paper is dedicated to the memory of Berislav Bu\v{c}a.}
\endgroup

Symmetries play a key role in understanding the dynamics of open quantum systems. Strong dynamical symmetries, first introduced by B. Bu\v{c}a \textit{et al.}~\cite{buca19}, explain how persistent oscillations arise in systems governed by the Lindblad equation. Specifically, a strong dynamical symmetry is an operator $A$ satisfying
\begin{equation}
    [H,A]=\omega A, \qquad [L_{j}, A] = [L_{j}^{\dagger},A] = 0 \qquad \forall j,
    \label{eq:SDS}
\end{equation}
i.e.\ it is a ladder operator of the Hamiltonian $H$ which commutes with every jump operator $L_{j}$ and its adjoint. Acting with powers of $A$ and $A^{\dagger}$ on the stationary state $\rho_{\rm ss}$ produces coherent eigenmodes that oscillate at frequencies that are integer multiples of $\omega$ and never decay, allowing the system to retain memory of its initial condition despite the dissipation. For finite-dimensional systems with a full-rank stationary state the converse also holds: persistent oscillations imply such a symmetry must exist~\cite{buca19,booker20}. The algebra generated by $A$, $A^\dagger$ and $\rho_{\rm ss}$ constructs the entire non-decaying sector and the resulting eigenmodes can lead to phenomena such as quantum synchronisation~\cite{tindall20,buca22} and time-crystalline order~\cite{buca19b,medenjak20}.

Persistent oscillatory behaviour has also been observed to emerge in infinite-dimensional dissipative systems involving collective spins: a dissipative analogue of a time crystal~\cite{wilczek12,sacha18,zaletel23} that has been realised in atom--cavity experiments~\cite{kessler21,kongkhambut22}. Examples include squeezed and strongly driven superradiance~\cite{sanchez19,nemeth25,nemeth26,liu25}, ensembles coupled through shared or cascaded reservoirs~\cite{stannigel12,pocklington24,cabot24,chu26,hajdusek22}, and interacting lattices~\cite{alaeian22}.

One of the simplest examples of this emergent oscillatory phenomenon is a driven ensemble of $N$ two-level atoms decaying through a common channel, often referred to as a `boundary time crystal'~\cite{iemini18,tucker18,piccitto21}. Its oscillatory phase has been proposed as a resource for sensing~\cite{montenegro23,cabot24}. This system has a unique stationary state at every finite $N$, with all other modes decaying in time~\cite{iemini18,booker20,minganti18}. As the spin $S=N/2\to\infty$, however, its magnetisation oscillates for increasingly longer times, at a frequency fixed by both the drive and the dissipation~\cite{iemini18,carollo22}. The stationary state is known exactly~\cite{puri79}, and semiclassical analyses describe the limiting trajectories and the slow approach to them~\cite{drummond78,carmichael80,drummond80,carollo22,nakanishi23,dutta25}. The system does not, however, have an operator satisfying the conditions of Eq.~\eqref{eq:SDS}, even as $S \rightarrow \infty$ and so appears to sit outside the strong dynamical symmetry paradigm.  Quantum treatments so far give the Liouvillian spectrum perturbatively in the dissipation strength~\cite{nemeth25} or characterise its eigenmodes numerically~\cite{nemeth26}. A quantum, algebraic construction of the system's emergent oscillatory modes: i.e. explicit quantum operators with controlled finite-size corrections that construct its non-decaying sector throughout the oscillatory regime as $S \rightarrow \infty$, is missing.

In this work we provide this construction. We take a collective spin in a linear field, damped by a general linear jump operator, and determine exactly the conditions for a ladder of oscillatory eigenmodes to emerge as $S \rightarrow \infty$. We then write down a closed-form expression for the `symmetry operator' $A_S$ which can be used, in tandem with the stationary state, to construct imaginary eigenmodes with residuals that vanish as $O(S^{-1})$. Together with a second operator $B_{S}$, an emergent conserved quantity, every emergent non-decaying mode with a smooth classical limit can be constructed. The slow relaxation of $B_S$ describes how quantum noise selects the stationary state. The driven, collectively decaying ensemble of Ref.~\cite{iemini18} is contained within our results, and thus we provide an explicit, quantum algebraic account of its emergent oscillatory sector.

Finally, we show that the same symmetry operators can also describe emergent oscillatory phenomena in a pair of interacting collective spins. Specifically, the stationary state of a single spin can be purified into the entangled dark state of two interacting spins sharing a reservoir~\cite{stannigel12,chu26}, and the operators $A_S$ and $B_S$, acting on the first spin, can be applied to this dark state to generate its oscillatory sector as $S \rightarrow \infty$. Quenching the interaction between the two spins then sets their entanglement oscillating, and we calculate an analytical expression for the time-dependent logarithmic negativity.

\begin{figure}[t]
\includegraphics[width=\columnwidth]{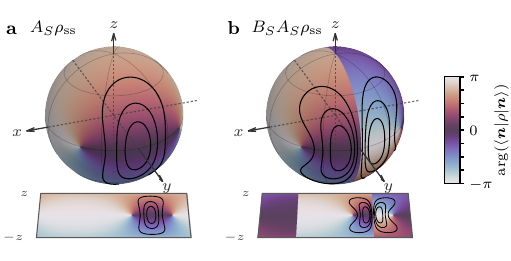}
\caption{Two emergent imaginary eigenmodes of the Liouvillian in Eq.~\eqref{eq:model}, $A_S\rho_{\rm ss}$ (a) and $B_SA_S\rho_{\rm ss}$ (b), with eigenvalue $-{\rm i}\Omega$, plotted on the Bloch sphere in the limit $S\to\infty$, for $\omega=3\kappa/2$. The phase ${\rm arg}(\langle \bm n |\rho \vert \bm n  \rangle)$ of each mode $\rho$ is given by the color, where $|\bm n\rangle=e^{-{\rm i}\varphi S^z}e^{-{\rm i}\vartheta S^y}|S,S\rangle$ is the spin coherent state along $\bm n=(\sin\vartheta\cos\varphi,\sin\vartheta\sin\varphi,\cos\vartheta)$. Some lines of constant modulus $ \vert \langle  \bm n |\rho \vert \bm n \rangle \vert$ are shown by the contours, which are at $25\%$, $50\%$ and $75\%$ of the maximum value. The factor $B_S$ changes sign across the great circle $n_x=0$. The phase is undefined at the cusps $\bm n_\pm=(\pm\Omega/\omega,\kappa/\omega,0)$, where $\langle\bm n|\rho|\bm n\rangle$ vanishes. The surface of each sphere is unrolled into a map below it.}
\label{fig:sphere}
\end{figure}

\begin{figure*}[t]
\includegraphics[width=\textwidth]{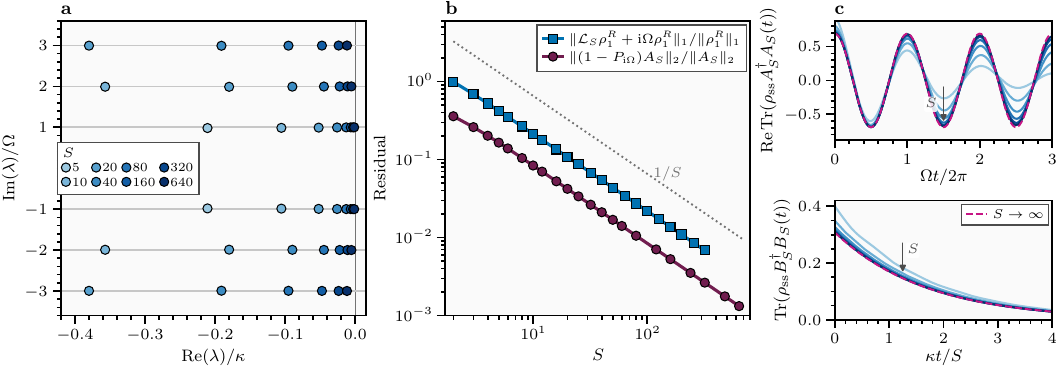}
\caption{Emergence of the quantum algebra for the Liouvillian in Eq.~\eqref{eq:model} with $\omega=2\kappa$, for $S$ ranging from $5$ to $640$ (light to dark). \textbf{a}~Least damped eigenvalue of the exact finite-$S$ Liouvillian $\Li_S$ near each harmonic ${\rm i}n\Omega$, $n = \pm 1, \pm2, \pm3$. \textbf{b}~Squares show the decaying relative trace-norm residual of the right mode $\rho^{R}_{1} = A_S\rho_{\rm ss}$ in Eq.~\eqref{eq:rightmodes}.
Circles show the relative weight, in Hilbert--Schmidt norm $\|\cdot\|_2$, of the left mode $A_S$ outside the eigenspace of $\Li_S^\dagger$ for the eigenvalues within $\omega/4$ of ${\rm i}\Omega$, with projector $P_{{\rm i}\Omega}$. The dotted line shows perfect $1/S$ decay for reference. \textbf{c}~Stationary two-time correlations ${\rm Re}\,{\rm Tr}\bigl(\rho_{\rm ss}A_S^\dagger A_S(t)\bigr)$ against $\Omega t / 2\pi$ (top) and ${\rm Tr}\bigl(\rho_{\rm ss}B_S^\dagger B_S(t)\bigr)$ against $\kappa t/S$ (bottom) for $S$ from $5$ to $160$. Dashed: the $S\to\infty$ limits, Eq.~\eqref{eq:correlation} (top) and the sum over the modes of Eq.~\eqref{eq:diffusion} at fixed $\kappa t/S$ (bottom).}
\label{fig:one}
\end{figure*}

\textit{A single collective spin.}---We start with a spin of length $S$ with operators $\bm S=(S^{x},S^{y},S^{z})$, evolving under a linear Hamiltonian $H = \bm h\cdot\bm S$ and a single nonunital linear jump
\begin{equation}
\Li_S\rho=-{\rm i}[\bm h\cdot\bm S,\rho]+\D[L_S]\rho,\qquad
L_S=\sqrt{\frac{\kappa}{S}}\,\bigl(\alpha S^{-}+\beta S^{+}\bigr),
\label{eq:model}
\end{equation}
with $\D[L]\rho=L\rho L^\dagger-\tfrac12\{L^\dagger L,\rho\}$, $|\alpha|^2-|\beta|^2=1$ and $S^{\pm}=S^{x}\pm {\rm i}S^{y}$. Here $\kappa > 0$. The field $\bm h=(h_{x},h_{y},h_{z})$, with $\omega=|\bm h|$, and the constants $\kappa,\alpha,\beta$ are independent of $S$. The $\sqrt S$ normalisation ensures the Hamiltonian and dissipative terms compete on an equal footing as $S$ grows. The driven, collectively decaying ensemble, introduced as a `boundary time crystal' in Ref.~\cite{iemini18}, corresponds to the case $\bm h=\omega\hat{\bm x}$ and $\beta=0$.

For every finite $S$ the stationary state of Eq.~\eqref{eq:model} is unique. For $h_{z}=0$ it is known in closed form~\cite{puri79,lourenco22}. The simplest derivation comes from noticing that adding a constant to a jump operator $L \rightarrow L + c$ is equivalent to the transformation $H \rightarrow H + H_c$ in the Hamiltonian, with $H_c=\tfrac{{\rm i}}{2}(c^*L-cL^\dagger)$. When $h_z=0$ the Hamiltonian is exactly of the form $H_c$ and can thus be brought entirely into the dissipator, so the whole Liouvillian is $\Li_S=\D[J]$ with
\begin{equation}
J=\sqrt{\frac{\kappa}{S}}\,\bigl(\alpha S^-+\beta S^+ +\gamma\bigr),
\label{eq:shiftedjump}
\end{equation}
where
\begin{equation}
\gamma=\frac{{\rm i}S}{\kappa}\bigl(\alpha(h_x-{\rm i}h_y)-\beta(h_x+{\rm i}h_y)\bigr).
\end{equation}
Any invertible $J$ then leads to the stationary state
\begin{equation}
\rho_{\rm ss}=\frac{(J^\dagger J)^{-1}}{\Tr \left( \,(J^\dagger J)^{-1} \right)},
\label{eq:ss}
\end{equation}
which satisfies $\D[J]\rho_{\rm ss} = 0$
since $J\rho_{\rm ss}J^\dagger$ and $\tfrac12\{J^\dagger J,\rho_{\rm ss}\}$ are both proportional to the identity and cancel. When $\omega>\kappa$, $J$ is
invertible and $\|J^{-1}\|=O(S^{-1/2})$ (see the Supplemental Material, SM), where we use $\| \cdot \|$ throughout to mean the operator norm. For
$\beta=0$, nilpotency of $S^-$ ensures invertibility for any
nonzero drive.

We can prove that the conditions
\begin{equation}
h_z=0\qquad\text{and}\qquad \omega>\kappa,
\label{eq:criterion}
\end{equation}
are necessary and sufficient for $\Li_S$ to support, as $S \to \infty$, an emergent ladder of eigenmodes with eigenvalues $\lambda = {\rm i}n\Omega$, where $n$ is an integer and $\Omega=\sqrt{\omega^2-\kappa^2}$. Sufficiency is proved below by explicit construction of the eigenmodes. We will prove necessity afterwards.

If $h_z = 0$ we are free to rotate about $z$, and so we set $\bm h=\omega\hat{\bm x}$ without loss of generality. Now, writing $s^{i}=S^{i}/S$ for the normalised spin components, we identify the following two key operators
\begin{equation}
\begin{aligned}
A_S&=\bigl(\omega s^{y}-\kappa+{\rm i}\Omega s^{z}\bigr)\bigl(\omega-\kappa s^{y}\bigr)^{-1},\\
B_S&=\Omega s^{x}\bigl(\omega-\kappa s^{y}\bigr)^{-1}.
\end{aligned}
\label{eq:dressed}
\end{equation}
For $\kappa=0$ they reduce to $s^{y}+{\rm i}s^{z}$ and $s^{x}$, the rotating and conserved components of a spin precessing about the field. They are bounded rational functions of the spin, well defined because $\omega-\kappa s^{y}$ is positive definite when $\omega > \kappa$ and $A_S$ also satisfies $1\le\|A_S\|\le\frac{\omega+\kappa+\Omega}{\omega-\kappa}$. Substituting them into Eq.~\eqref{eq:model} it is possible to prove
\begin{equation}
\bigl\|\Li_S^\dagger A_S-{\rm i}\Omega A_S\bigr\|\le\frac{C}{S},\quad \bigl\|\Li_S^\dagger B_S\bigr\|\le\frac{C}{S},
\label{eq:residual}
\end{equation}
with $C$ independent of $S$ (see SM). Equation~\eqref{eq:residual} is the asymptotic form of the relationships induced by a strong-dynamical-symmetry $\Li^\dagger A={\rm i}\Omega A$ and by a conserved quantity $\Li^\dagger B=0$. Because the residuals are small in operator norm, they propagate through products allowing one to prove that
\begin{equation}
\Li_S^\dagger\rho^{L}_{n}={\rm i}n\Omega\,\rho^{L}_{n}+O(S^{-1}),\qquad \rho^{L}_{n}=f(B_S)A_S^{\,n},
\label{eq:general}
\end{equation}
for $n=0,1,2,\dots$ and any function $f(x)$ analytic on $[-1,1]$, the interval containing the spectrum of $B_S$. The $O(S^{-1})$ remainders here and below are bounded in operator norm for left modes and in trace norm for right modes, with constants that do not depend on $S$ but grow as $\omega$ approaches $\kappa$. The mode $f(B_S)(A_S^\dagger)^n$ has frequency $-n\Omega$.
These are the left eigenmodes of the Liouvillian as $S \rightarrow \infty$, in the sense that their residuals vanish and, as Fig.~\ref{fig:one}(a) shows, the least-damped eigenvalues of the exact $\Li_S$ converge to ${\rm i}n\Omega$.  Acting on the stationary state, which is well defined throughout this oscillatory regime, gives the corresponding right modes,
\begin{equation}
\Li_S\rho^{R}_{n}=-{\rm i}n\Omega\,\rho^{R}_{n}+O(S^{-1}),\qquad \rho^{R}_{n}=f(B_S)A_S^{\,n}\rho_{\rm ss}.
\label{eq:rightmodes}
\end{equation}
Products such as $A_S^{\,n}\rho_{\rm ss}(A_S^\dagger)^{m}$, which directly appear in the presence of a strong dynamical symmetry~\cite{buca19}, are contained within this form of the right modes, since $A_S^\dagger A_S=1-B_S^2+O(S^{-1})$ and $\rho_{\rm ss}$ commutes with $A_S$ up to $O(S^{-1})$.
Because the evolution is a contraction, the residual in Eq.~\eqref{eq:rightmodes} controls the dynamics directly, $e^{t\Li_S}\rho^R_{n}=e^{-{\rm i}n\Omega t}\rho^R_{n}+O(\kappa t/S)$, so these modes oscillate without decay on timescales $t\ll S/\kappa$.

Figure~\ref{fig:sphere} shows two of the right eigenmodes, $A_S\rho_{\rm ss}$ and $B_SA_S\rho_{\rm ss}$, in the limit $S\to\infty$ on the Bloch sphere. Figure~\ref{fig:one}(a), meanwhile, shows the imaginary eigenvalues emerging from the exact finite-$S$ spectrum of the Liouvillian. Figure~\ref{fig:one}(b) shows the relevant residuals vanishing as $1/S$.

If the conditions in Eq.~\eqref{eq:criterion} fail, no operator with a nonzero, continuous classical limit can oscillate indefinitely at a fixed nonzero frequency as $S \rightarrow \infty$. When they hold, any such operator with a smooth classical limit that oscillates is approximated arbitrarily well by the operators of Eq.~\eqref{eq:general}, so the construction is complete for operators with smooth classical limits. We prove this in the SM.

The operators in Eq.~\eqref{eq:dressed} originate from the appropriate quantisation of two key classical variables in the limit $S\to\infty$. In this limit a spin prepared in a coherent state can be viewed as a classical unit vector $\bm n(t)={\rm Tr}(\rho(t)\,\bm S)/S$~\cite{arecchi72,lieb73} which precesses about the field while the dissipation pulls it towards the south pole, $\dot{\bm n}=\bm h\times\bm n-\kappa(\hat{\bm z}-n_z\bm n)$~\cite{drummond80,carollo22}. When $\omega>\kappa$, the two effects balance on closed orbits: every trajectory apart from two fixed points is periodic with the same frequency $\Omega$. The natural coordinates are then a conserved component and a rotating one,
\begin{equation}
\xi=\frac{\Omega n_x}{\omega-\kappa n_y},\qquad q=\frac{\omega n_y-\kappa+ {\rm i}\Omega n_z}{\omega-\kappa n_y},
\label{eq:dressedsphere}
\end{equation}
with $\xi^2+|q|^2=1$. In these coordinates the motion is a rigid rotation: $\xi$ is constant and $q$ simply rotates, $q(t)=q(0)e^{{\rm i}\Omega t}$. The variables $\xi$ and $q$ can be mapped directly onto the orbit label and phase used by Drummond and Carmichael~\cite{drummond78,carmichael80}, who studied the case $\beta=0$. The operators $A_S$ and $B_S$ in Eq.~\eqref{eq:dressed} are quantisations of $q$ and $\xi$. The factor $(\omega-\kappa s^y)^{-1}$ turns the damped precession into a rigid rotation. A spin prepared in the coherent state $\rho(0)=|\bm n_0\rangle\langle\bm n_0|$, for instance $|S,-S\rangle$, has ${\rm Tr}\bigl(\rho(t)A_S\bigr)\to q(\bm n_0)\,e^{{\rm i}\Omega t}$ as $S\to\infty$ at fixed $t$.

As $S\to\infty$ the system has a continuum of stationary modes, $f(B_S)\rho_{\rm ss}$ for any such $f$. This is the infinite-dimensional analogue of how each symmetry sector of a strong symmetry --- first introduced by Bu\v{c}a and Prosen \cite{buca12} --- identifies a unique stationary state.

In the classical variable picture, $\xi$ is conserved as $S \rightarrow \infty$ and so any distribution over $\xi$ is stationary. At finite $S$, however, the quantum noise, of order $1/S$, lifts the continuous degeneracy. It acts over times of order $S/\kappa$ and so it selects the stationary state when taking $t \rightarrow \infty$ and then $S \rightarrow \infty$. For the `boundary time crystal' case $\beta=0$ the selected state has the distribution $p(x)\propto1/(\Omega^2+\kappa^2x^2)$ over the values of $B_S$~\cite{drummond78,carmichael80}, in the sense that ${\rm Tr}\bigl(\rho_{\rm ss}f(B_S)\bigr)\to\int_{-1}^{1}p(x)f(x)\,dx$ as $S\to\infty$. Here $B_S$ is Hermitian up to $O(S^{-1})$, so these are the statistics of a measurement of $B_S$.

The distribution also sets the two-time correlations. In the stationary state ${\rm Tr}(\rho_{\rm ss}A_S)\to0$ as $S \to \infty$, but two-time correlations don't decay. Taking $A_S(t)=e^{t\Li_S^\dagger}A_S$, we can show
\begin{equation}
\lim_{S\to\infty}{\rm Tr}\bigl(\rho_{\rm ss}A_S^\dagger A_S(t)\bigr)=\Bigl(\frac{\omega^2}{\kappa^2}-\frac{\Omega}{\kappa\theta}\Bigr)e^{{\rm i}\Omega t},
\label{eq:correlation}
\end{equation}
where $\theta=\arctan(\kappa/\Omega)$ and which we observe directly in Fig.~\ref{fig:one}(c). The frequency is predicted by the classical rotation, but the amplitude $\int_{-1}^{1}p(x)(1-x^2)\,dx$ comes from the distribution that the quantum noise selected.

We can also calculate how quickly this selection happens. The operators $f_k(B_S)$ relax at rates $\kappa\mu_k/S$ up to $O(S^{-1})$ corrections, and numerically the slowest real eigenvalue of $\Li_S$ approaches $-\kappa\mu_1/S$. Here $f_k$ and $\mu_k$ are the solutions of the operator form of the orbit Fokker--Planck equation of Refs.~\cite{drummond78,carmichael80},
\begin{equation}
\frac{1}{4\Omega^2}\bigl(\Omega^2+\kappa^2x^2\bigr)\frac{d}{dx}\Bigl[(1-x^2)\,f_k'(x)\Bigr]=-\mu_k\,f_k(x)
\label{eq:diffusion}
\end{equation}
on $-1\le x\le1$. The lowest solution, $\mu_0=0$ with $f_0(x)=1$, corresponds to the stationary state, and the next one, $\mu_1$, sets the slowest relaxation. It depends only on $\kappa/\omega$ and tends to $1/2$ for $\kappa\ll\omega$, so in that regime the stationary state is selected over times of order $S/\kappa$. The correlation of $B_S$ shows this directly. It is, to excellent accuracy, a single exponential, ${\rm Tr}\bigl(\rho_{\rm ss}B_S^\dagger B_S(t)\bigr)\simeq\bigl(\tfrac{\Omega}{\kappa\theta}-\tfrac{\Omega^2}{\kappa^2}\bigr)e^{-\kappa\mu_1t/S}$, with an amplitude equal to one minus that of Eq.~\eqref{eq:correlation}. Figure~\ref{fig:one}(c) shows the exact finite-$S$ correlation of $A_S$ converging to Eq.~\eqref{eq:correlation}, and the correlation of $B_S$ relaxing as Eq.~\eqref{eq:diffusion} predicts.

\textit{Two interacting collective spins.}---The single-spin stationary state Eq.~\eqref{eq:ss} can be obtained from a partial trace over a pure state of two spins. Since $\rho_{\rm ss}\propto J^{-1}(J^{-1})^\dagger$, such a purification is $|\Psi\rangle \propto (J^{-1}\otimes I)|\Phi\rangle$ for any maximally entangled state $|\Phi\rangle$ of the two spins. We now show that, for the right choice of $|\Phi\rangle$, this pure state is the exact stationary state of a pair of interacting spins sharing a reservoir, and that the whole quantum algebra derived above survives in this system. Take two spins of length $S$, with operators $\bm S_1$ and $\bm S_2$, driven by the same field and coupled to a common lowering channel,
\begin{equation}
\begin{aligned}
\Li_S^{(2)}\rho&=-{\rm i}[H,\rho]+\frac{\Gamma}{S}\D[S_1^{-}+S_2^{-}]\rho,\\
H&=\omega(S_1^x+S_2^x)+\frac{g}{S}\bigl(S_1^xS_2^y-S_1^yS_2^x\bigr),
\end{aligned}
\label{eq:pair}
\end{equation}
with $\Gamma>0$ the rate of the shared decay channel and $\omega>g>0$. The Liouvillian in Eq.~\eqref{eq:pair} describes two ensembles coupled through a shared waveguide, the setting of cascaded and chiral quantum optics~\cite{gardiner93,carmichael93,lodahl17}. For unidirectional coupling, $g=\Gamma$: the first ensemble drives the second without receiving feedback~\cite{stannigel12,cabot24,chu26}. For $g\neq\Gamma$, both ensembles influence each other.

Equation~\eqref{eq:pair} has a pure stationary state $\rho^{(2)}_{\rm ss} = |\Psi_S\rangle \langle \Psi_S \vert$ for every $S$ and every $\Gamma$ with
\begin{equation}
|\Psi_S\rangle=\sqrt{\tfrac{2S+1}{Z}}\,\bigl(T_S^{-1}\otimes I\bigr)|s_0\rangle,\qquad T_S=s^{-}+{\rm i}\frac{\omega}{g},
\label{eq:dark}
\end{equation}
where $|s_0\rangle=\sum_m(-1)^{S-m}|m,-m\rangle/\sqrt{2S+1}$ is the two-spin singlet, the unique state of total spin zero, and $s^{-}=\frac{S^-}{S}$. The scale factor $Z$ ensures a trace of unity. Tracing out the second spin gives the single-spin stationary state in Eq.~\eqref{eq:ss}, with $g$ replacing $\kappa$ in Eq.~(\ref{eq:shiftedjump}). The state is not symmetric under exchange of the spins, reflecting the chiral interaction. Since both the collective jump and the Hamiltonian annihilate $|\Psi_S\rangle$, it is an exact stationary state~\cite{stannigel12,chu26}, entangled across the two spins with full Schmidt rank.

\begin{figure*}[t]
\includegraphics[width=\textwidth]{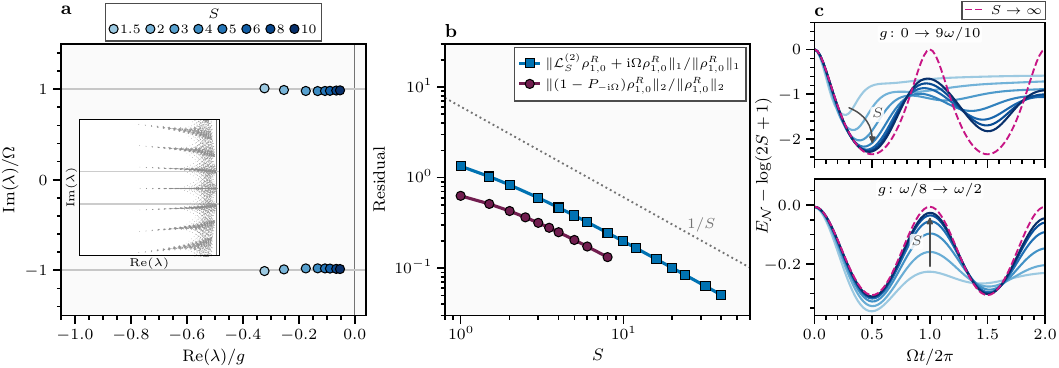}
\caption{Emergence of the quantum algebra for the two spin Liouvillian $\Li^{(2)}_S$ in Eq.~\eqref{eq:pair} with $g=\omega/2$ and $\Gamma=g/2$ in panels a and b, and $\Gamma=\omega/2$ in panel c. \textbf{a}~Least damped eigenvalue of $\Li^{(2)}_S$ in the rows $\mathrm{Im}\,\lambda\simeq\pm\Omega$, $\Omega=\sqrt{\omega^2-g^2}$, for increasing $S$ (light to dark). Inset shows the full spectrum for $S = 5$. \textbf{b}~Squares show the relative trace-norm residual of the right mode $\rho^{R}_{1,0}=A_1\rho^{(2)}_{\rm ss}$ in Eq.~\eqref{eq:pairmodes}. Circles show its relative weight, in Hilbert--Schmidt norm $\|\cdot\|_2$, outside the eigenspace of $\Li^{(2)}_S$ for the eigenvalues within $\omega/4$ of $-{\rm i}\Omega$, with projector $P_{-{\rm i}\Omega}$. \textbf{c}~Logarithmic negativity minus $\log(2S+1)$ after a quench of the interaction $g'\rightarrow g$ from the stationary state at $g'$ (colours, increasing $S$). Dashed: the $S\to\infty$ result given by Eq.~\eqref{eq:ENt} (top) and the general $g'\neq0$ case stated in the SM (bottom).}
\label{fig:two}
\end{figure*}

We can identify and construct emergent imaginary eigenmodes of this system using $A_S$ and $B_S$, the single-spin operators defined in Eq.~\eqref{eq:dressed} with $\kappa\to g$. First, we write $A_1=A_S\otimes I$ and $B_1=B_S\otimes I$ for their action on the first spin. We can then prove (see SM) the following
\begin{equation}
\begin{aligned}
\Li_S^{(2)}\rho^{R}_{n,m}&=-{\rm i}(n-m)\Omega\,\rho^{R}_{n,m}+O(S^{-1}),\\
\rho^{R}_{n,m}&=f(B_1)A_1^{\,n}\rho^{(2)}_{\rm ss}(A_1^\dagger)^{m}g(B_1)^\dagger,
\end{aligned}
\label{eq:pairmodes}
\end{equation}
for any $f$ and $g$ analytic on $[-1,1]$, thus identifying emergent imaginary eigenmodes in the limit $S \rightarrow \infty$. Here $\Omega=\sqrt{\omega^2-g^2}$ is set by the drive and the interaction, independent of $\Gamma$. The bath rate enters only at order $1/S$, through the rate at which these modes decay.

\textit{Entanglement.}---Because $\rho_{\rm ss}^{(2)}$ is pure, its entanglement across the two spins is fixed by its Schmidt spectrum, and the squared Schmidt coefficients are the eigenvalues of the single-spin state in Eq.~\eqref{eq:ss}. More generally, for any excitation $(O\otimes I)|\Psi_S\rangle$ the Schmidt coefficients are the singular values of the single-spin matrix $M=O\,T_S^{-1}$, which reduces every two-spin entanglement measure of these states to a single-spin calculation.

Here we give the thermodynamic limit of the negativity of $\rho^{(2)}_{\rm ss}$ in closed form. A trace expansion, carried out in the SM, gives the logarithmic negativity $\EN(\rho)=\log\|\rho^{T_2}\|_1$~\cite{vidal02,plenio05}, where $T_2$ is the partial transpose on the second spin and $\|\cdot\|_1$ the trace norm,
\begin{equation}
\EN(\rho^{(2)}_{\rm ss})=\log(2S+1)+\log\left(\frac{\Omega\,\theta}{g}\right)+O(S^{-1}),
\label{eq:ENss}
\end{equation}
with $\theta=\arctan(g/\Omega)$ as before, now with $g$ in place of $\kappa$. The leading term is that of a maximally entangled state of two $(2S+1)$-level systems. The deficit is finite, negative, and fixed by the ratio $g/\omega$ alone: it vanishes as $g/\omega\to0$, where $|\Psi_S\rangle$ becomes the singlet, and grows as the interaction approaches the drive. Dark steady states of spin ensembles sharing a reservoir have previously been constructed in Ref.~\cite{chu26}, where their metrological entanglement was characterised.

Finally, to observe the role of the quantum algebra in the entanglement dynamics we consider a quench. The pair starts in its stationary state $|\Psi'_S\rangle$ at interaction strength $0\le g'<\omega$, with $g'=0$ the singlet, and the interaction is then switched to $g$ and the system evolved in time. Both dark states (at $g$ and $g'$) descend from the same singlet, so the pure initial state can be written as an excitation of the new stationary state, $|\Psi'_S\rangle\propto(O\otimes I)|\Psi_S\rangle$ with $O=T'^{-1}_ST_S$ and $T'_S=s^{-}+{\rm i}\omega/g'$. The excitation $O$ is a combination of the operators $f(B_S)A_S^{\,n}$ and $f(B_S)(A_S^\dagger)^n$, each of which rotates at frequency $n\Omega$ under the new coupling.
The same trace expansion as we used for the steady state then gives $\EN(\rho(t))$ for every fixed $t$ as $S\to\infty$, as an explicit double integral for a general quench $g'\to g$ (see SM). For the quench from the singlet, $g'=0$, it takes the simple closed form
\begin{equation}
\EN(\rho(t))=\log(2S+1)+2\log\left(\frac{\sinh^{-1}(u_t)}{u_t\sqrt{1+u_t^2}}\right)+O(S^{-1}),
\label{eq:ENt}
\end{equation}
where $
u_t=\frac{g}{\Omega}\Bigl|\sin\frac{\Omega t}{2}\Bigr|$.
Figure~\ref{fig:two}(c) shows the exact evolution approaching Eq.~\eqref{eq:ENt} as $S$ increases, together with a quench from $g'\neq0$ approaching the general double integral result of the SM.

\textit{Conclusion.}---Strong dynamical symmetries explain persistent oscillations in finite dimensional open quantum systems via the existence of an operator satisfying Eq.~(\ref{eq:SDS}). We have shown that for collective spins there is a bounded rational function $A_{S}$ of the spin operators which behaves like an emergent counterpart to this operator up to corrections of order $1/S$. Together with a second operator, $B_S$, an emergent conserved quantity, it generates the non-decaying modes which become exact eigenmodes as $S \rightarrow \infty$. The slow relaxation of $B_S$ sets how quantum noise selects the stationary state. We then showed how the same operators generate the oscillatory sector of a pair of interacting spins, allowing us to derive the entanglement dynamics after a quench.

The construction in this paper can be extended to a many-body setting by coupling neighboring pairs of spins through their collective jumps, $H_{\rm int}=(\chi/S)\sum_{b=1}^{n}(C_b^\dagger C_{b+1}+{\rm h.c.})$ with $C_b=S_{b1}^-+S_{b2}^-$. Open questions include calculating the closed form entanglement dynamics of such a many-body system through the resulting quantum algebra and understanding the interplay between the limits $n \rightarrow \infty$ and $S \rightarrow \infty$.

\begin{acknowledgments}
\textit{Acknowledgments.}---This paper is dedicated to the memory of Berislav Bu\v{c}a. His deep intellect and creativity are responsible for much of our understanding about non-stationarity in open quantum systems and had a profound influence on my early years as a physicist.

The author is grateful for ongoing support through the Flatiron Institute, a division of the Simons Foundation, and for discussions with Carlos Sánchez Muñoz. Frontier large language models were used during the development of this work.
\end{acknowledgments}

\bibliography{main}
\end{document}


\renewcommand{\theequation}{S\arabic{equation}}
\title{Supplemental Material: Emergent quantum algebras in dissipative time crystals}
\author{Joseph Tindall}
\affiliation{Center for Computational Quantum Physics, The Flatiron Institute, 162 5th Avenue, New York, New York 10010, USA}
\date{\today}
\maketitle

\section{Notation}
\label{sec:tools}
Here, equation numbers without an S prefix refer to equations which are restated verbatim from the main text with the identical equation number. We write $d=2S+1$ for the dimension of a spin of length $S$ and $\bm s=\bm S/S$ for its normalised spin operators. Each component $s^i$ has eigenvalues $m/S$ with $m=-S,\dots,S$, so its operator norm, its largest singular value, is one.

\textit{The Liouvillian.} The single spin evolves under $\dot\rho=\Li_S\rho$, with $\Li_S\rho=-\ii[\bm h\cdot\bm S,\rho]+\sum_\mu\D[L_\mu]\rho$ for several jumps and $\D[L]\rho=L\rho L^\dagger-\tfrac12\{L^\dagger L,\rho\}$. The main text has a single jump,
\begin{equation}
\begin{aligned}
\Li_S\rho&=-\ii[\bm h\cdot\bm S,\rho]+\D[L_S]\rho,\\
L_S&=\sqrt{\kappa/S}\,\bigl(\alpha S^{-}+\beta S^{+}\bigr),
\end{aligned}\tag{\mnum{eq:model}}
\end{equation}
with $|\alpha|^2-|\beta|^2=1$, but several of the results below hold for any finite number of linear jumps and we state them under that generality. Some results are stated only for $\beta=0$, where we can drop $\alpha$ and just take $L_S=\sqrt{\kappa/S}\,S^-$. This is the `boundary time crystal' case of the main text.

Operators evolve under the adjoint equation, $\dot X=\Li_S^\dagger X$, with
\begin{equation}
\Li_S^\dagger X=\ii[\bm h\cdot\bm S,X]+\sum_\mu\Bigl(L_\mu^\dagger XL_\mu-\tfrac12\{L_\mu^\dagger L_\mu,X\}\Bigr),
\label{eq:Ldag}
\end{equation}
so that ${\rm Tr}\bigl(\rho(t)X\bigr)={\rm Tr}\bigl(\rho(0)\,e^{t\Li_S^\dagger}X\bigr)$. A left eigenmode is an operator $X$ with $\Li_S^\dagger X=\lambda^*X$ and a right eigenmode is an operator $\rho$ with $\Li_S\rho=\lambda\rho$. Both evolutions are contractions, $\|e^{t\Li_S^\dagger}X\|\le\|X\|$ and $\|e^{t\Li_S}\rho\|_1\le\|\rho\|_1$, where $\|X\|=\max_{\|\psi\|=1}\|X\psi\|$ is the operator norm, equal to the largest singular value of $X$, and $\|X\|_1=\Tr\bigl(\sqrt{X^\dagger X}\bigr)$ is the trace norm. This contractive property is what lets small residuals control the dynamics.

\textit{Coherent states and symbols.} Spin coherent states connect operators to functions on the sphere~\cite{arecchi72,lieb73}. For a point $\bm n=(\sin\vartheta\cos\varphi,\sin\vartheta\sin\varphi,\cos\vartheta)$ on the unit sphere, $|\bm n\rangle=e^{-\ii\varphi S^z}e^{-\ii\vartheta S^y}|S,S\rangle$ is the state with the spin pointing along $\bm n$. As $S\to\infty$ the expectation value $\langle\bm n|X|\bm n\rangle$ of a spin operator $X$ tends to a function $\hat X(\bm n)$, which we call the symbol of $X$. For example, $\bm s$ has symbol $\bm n$ and $s^{-}$ has symbol $n_x-\ii n_y$. The key operators of the main text,
\begin{equation}
\begin{aligned}
A_S&=\bigl(\omega s^{y}-\kappa+\ii\Omega s^{z}\bigr)\bigl(\omega-\kappa s^{y}\bigr)^{-1},\\
B_S&=\Omega s^{x}\bigl(\omega-\kappa s^{y}\bigr)^{-1},
\end{aligned}\tag{\mnum{eq:dressed}}
\end{equation}
have symbols $\hat A_S=q$ and $\hat B_S=\xi$, where
\begin{equation}
\xi=\frac{\Omega n_x}{\omega-\kappa n_y},\qquad q=\frac{\omega n_y-\kappa+\ii\Omega n_z}{\omega-\kappa n_y}\tag{\mnum{eq:dressedsphere}}
\end{equation}
are the classical coordinates of the main text. We use $\xi$ for this function on the sphere and, when needed, $x\in[-1,1]$ for a value it takes, as the argument of one-variable functions such as $f(x)$ and $p(x)$ and as an integration variable. We also write $\overline{a}=\frac1{4\pi}\int a(\bm n)\,d\bm n$ for the average of a function over the sphere, and $\max|a|=\max_{\bm n}|a(\bm n)|$ for its largest absolute value.

In the other direction, a function $a(\bm n)$ on the sphere defines the operator
\begin{equation}
\Op_S(a)=\frac{d}{4\pi}\int a(\bm n)\,|\bm n\rangle\langle\bm n|\,d\bm n,
\end{equation}
which has trace $\Tr\bigl(\Op_S(a)\bigr)=d\,\overline{a}$ and operator norm $\|\Op_S(a)\|\le\max|a|$. Products of these operators follow products of their functions,
\begin{equation}
\Op_S(a)\Op_S(b)=\Op_S(ab)+O(S^{-1}),
\label{eq:product}
\end{equation}
for smooth $a$ and $b$, with a constant set by finitely many derivatives of $a$ and $b$~\cite{bordemann94}. Any fixed polynomial in $\bm s$ equals $\Op_S$ of its symbol up to $O(S^{-1})$. We say that an operator $X_S$ has a continuous classical limit $y$ if $X_S=\Op_S(y)+o(1)$ in norm for a continuous function $y$, and a smooth classical limit if $y$ is smooth.

The Poisson bracket of two functions on the sphere is
\begin{equation}
\{a,b\}(\bm n)=\bm n\cdot\bigl(\nabla a\times\nabla b\bigr),
\label{eq:poisson}
\end{equation}
where $\nabla$ is the gradient with respect to $\bm n$. Only the components of $\nabla a$ and $\nabla b$ tangent to the sphere contribute, so $\{a,b\}$ depends only on the values of $a$ and $b$ on the sphere. For the coordinates themselves $\{n_i,n_j\}=\epsilon_{ijk}n_k$, the classical version of $[s^i,s^j]=\frac{\ii}{S}\epsilon_{ijk}s^k$. More generally, for fixed polynomials $X$ and $Y$ in $\bm s$,
\begin{equation}
[X,Y]=\frac{\ii}{S}\Op_S\bigl(\{\hat X,\hat Y\}\bigr)+O(S^{-2}).
\label{eq:commutator}
\end{equation}

\section{Oscillation criterion}
\label{sec:criterion}
The main text claims that the Liouvillian of Eq.~\mref{eq:model} supports, as $S\to\infty$, an emergent ladder of eigenmodes with eigenvalues ${\rm i}n\Omega$ if and only if
\begin{equation}
h_z=0\qquad\text{and}\qquad\omega>\kappa,\tag{\mnum{eq:criterion}}
\end{equation}
with $\Omega=\sqrt{\omega^2-\kappa^2}$. Here we prove this for any finite number of linear jumps $L_\mu=(\bm u_\mu+\ii\bm v_\mu)\cdot\bm S/\sqrt S$, where $\bm u_\mu$ and $\bm v_\mu$ are real-valued vectors. We first work out the classical dynamics, and then show that the quantum operators follow it.

\textit{The classical flow.} The jumps enter the classical dynamics only through their combined dissipative axis $\bm r=\sum_\mu\bm u_\mu\times\bm v_\mu$. To see this, apply Eq.~\eqref{eq:Ldag} to the normalised spin operators $s^i = S^i / S$, $i\in\{x,y,z\}$. The result is,
\begin{equation}
\Li_S^\dagger s^i=(\bm h\times\bm s)^i+r^i-\tfrac12\{\bm r\cdot\bm s,s^i\}+O(S^{-1}),
\label{eq:spinheis}
\end{equation}
where $\{\cdot,\cdot\}$ here is the anticommutator. Replacing $\bm s$ by its symbol $\bm n$ and dropping the $1/S$ terms gives the classical flow~\cite{drummond80,carollo22}
\begin{equation}
\dot{\bm n}=\bm h\times\bm n+\bm r-(\bm r\cdot\bm n)\,\bm n.
\label{eq:flow}
\end{equation}
The first term precesses $\bm n$ about the field. The other two pull $\bm n$ towards $\bm r/|\bm r|$ while keeping it on the unit sphere.

\textit{The flow as a linear equation.} Let $\bm\sigma=(\sigma^x,\sigma^y,\sigma^z)$ be the Pauli matrices, and write
\begin{equation}
\bm n=\frac{\langle\psi|\bm\sigma|\psi\rangle}{\langle\psi|\psi\rangle}
\end{equation}
for a two-component spinor $|\psi\rangle$. This is a one-to-one correspondence between points on the sphere and spinors taken up to an overall complex factor, which means the corresponding point on the sphere is determined uniquely by the ratio of the two components of $|\psi\rangle$. If $|\psi\rangle$ obeys the linear equation
\begin{equation}
\frac{d}{dt}|\psi\rangle=\Lambda|\psi\rangle,\qquad \Lambda=\tfrac12(\bm r-\ii\bm h)\cdot\bm\sigma,
\label{eq:spinor}
\end{equation}
then $\bm n$ obeys Eq.~\eqref{eq:flow}. The anti-Hermitian part of $\Lambda$ rotates the spinor, which precesses $\bm n$ about $\bm h$. The Hermitian part stretches the spinor along the eigenvector of $\bm r\cdot\bm\sigma$, which pulls $\bm n$ towards $\bm r$. So the nonlinear flow on the sphere is the image of the linear flow $|\psi(t)\rangle=e^{t\Lambda}|\psi(0)\rangle$, and everything about it is fixed by the two eigenvalues of $\Lambda$. These are $\pm\lambda$ with
\begin{equation}
\lambda^2=\tfrac14\bigl(|\bm r|^2-|\bm h|^2-2\ii\,\bm h\cdot\bm r\bigr).
\label{eq:lambda}
\end{equation}

Suppose $\lambda\neq0$, and let $|e_\pm\rangle$ be the two eigenvectors of $\Lambda$. A spinor proportional to $|e_+\rangle$ or $|e_-\rangle$ only changes its length and phase, so these give the two fixed points of the flow. Any other spinor is $|\psi\rangle=c_+|e_+\rangle+c_-|e_-\rangle$ with both amplitudes nonzero, and its point on the sphere is set by the ratio $c_+/c_-$, which evolves as $e^{2\lambda t}$. There are two cases.
\begin{itemize}
\item If $\lambda$ is purely imaginary, the ratio only rotates in phase. Every trajectory other than the two fixed points is a closed orbit, and all orbits share the frequency $\Omega=2|\lambda|$.
\item If $\lambda$ has a nonzero real part, the ratio grows or shrinks exponentially. Every trajectory other than the fixed points approaches one fixed point as $t\to\infty$ and the other as $t\to-\infty$.
\end{itemize}
If $\lambda=0$, which needs $\bm h\cdot\bm r=0$ and $|\bm h|=|\bm r|$, there is a single fixed point and every trajectory approaches it in both time directions. By Eq.~\eqref{eq:lambda}, $\lambda$ is purely imaginary and nonzero exactly when $\bm h\cdot\bm r=0$ and $|\bm h|>|\bm r|$. For the single jump of the main text this is Eq.~\mref{eq:criterion}, with $|\bm r|=\kappa$ and $\Omega=\sqrt{|\bm h|^2-|\bm r|^2}$.

\textit{Sufficiency.} Assume now that Eq.~\mref{eq:criterion} holds. We can take $\bm h=\omega\hat{\bm x}$ and $\bm r=-\kappa\hat{\bm z}$ without loss of generality. The coordinates $\xi$ and $q$ of Eq.~\mref{eq:dressedsphere} turn the flow into a rigid rotation, $\dot\xi=0$ and $\dot q=\ii\Omega q$, as can be checked directly from Eq.~\eqref{eq:flow}. The operators $A_S$ and $B_S$ of Eq.~\mref{eq:dressed} are the quantum versions of $q$ and $\xi$. We now show that they inherit this rotation up to corrections of order $1/S$, which allows us to prove the key bounds of the main text,
\begin{equation}
\bigl\|\Li_S^\dagger A_S-\ii\Omega A_S\bigr\|\le\frac{C}{S},\quad \bigl\|\Li_S^\dagger B_S\bigr\|\le\frac{C}{S}.\tag{\mnum{eq:residual}}
\end{equation}

First, split $A_S=ND^{-1}$ into its numerator $N=\omega s^y-\kappa+\ii\Omega s^z$ and denominator $D=\omega-\kappa s^y$. Both are linear in the spin, so Eq.~\eqref{eq:spinheis} gives their evolution,
\begin{equation}
\begin{aligned}
\Li_S^\dagger D&=\tfrac12\{V,D\}+O(S^{-1}),\\
\Li_S^\dagger N&=\ii\Omega N+\tfrac12\{V,N\}+O(S^{-1}),
\end{aligned}
\end{equation}
with $V=\kappa s^z$ and $\{\cdot,\cdot\}$ the anticommutator. These are the quantum versions of $\dot D=VD$ and $\dot N=(\ii\Omega+V)N$ along the classical flow. So $N$ rotates at frequency $\Omega$ while $D$ does not, and both grow or shrink at the same rate $V$, which cancels in the ratio $A_S=ND^{-1}$.

Applying the adjoint Liouvillian to products of operators gives,
\begin{equation}
\Li_S^\dagger(XY)=(\Li_S^\dagger X)Y+X(\Li_S^\dagger Y)+\mathcal P(X,Y),
\end{equation}
with $\mathcal P(X,Y)=\sum_\mu[L_\mu^\dagger,X][Y,L_\mu]$. Each commutator of a jump operator with a bounded function of $\bm s$ is of order $1/\sqrt S$. This is because a jump $L_\mu=(\bm u_\mu+\ii\bm v_\mu)\cdot\bm S/\sqrt S$ is $\sqrt S$ times a linear function of $\bm s$, while each commutator of normalised spins costs a factor $1/S$ by Eq.~\eqref{eq:commutator}. So $\mathcal P$ is of order $1/S$. The inverse $D^{-1}$ has to be treated separately, since it is not a polynomial in $\bm s$. It is bounded, $\|D^{-1}\|\le(\omega-\kappa)^{-1}$, because $D$ is positive definite with smallest eigenvalue at least $\omega-\kappa$. Its commutator with a jump follows from that of $D$, $[L_\mu,D^{-1}]=-D^{-1}[L_\mu,D]D^{-1}$, and so is also of order $1/\sqrt S$. Applying the product rule to $A_S=ND^{-1}$ and to $DD^{-1}=I$ then gives
\begin{equation}
\Li_S^\dagger A_S-\ii\Omega A_S=\tfrac12[V,A_S]+O(S^{-1})=O(S^{-1}),
\end{equation}
since $[V,A_S]$ is itself of order $1/S$. This proves Eq.~\mref{eq:residual} for $A_S$, with a constant $C$ that depends on the field and jump coefficients but not on $S$.

The same argument applies to $B_S=\Omega s^xD^{-1}$. Its numerator obeys $\Li_S^\dagger(\Omega s^x)=\tfrac12\{V,\Omega s^x\}+O(S^{-1})$, with no rotating term, so $\Li_S^\dagger B_S=O(S^{-1})$. It also applies to products. Since $\mathcal P(X, Y)$ between fixed powers of $A_S$, $A_S^\dagger$ and $B_S$ is of order $1/S$, induction on the number of factors shows that $f(B_S)A_S^n$ has residual of order $1/S$ for any polynomial $f$. The main text states the result for any $f$ analytic on $[-1,1]$, which needs one more step.

Such an $f$ extends analytically to a neighbourhood of $[-1,1]$, so Cauchy's formula writes $f(B_S)=\frac{1}{2\pi\ii}\oint f(z)\,(z-B_S)^{-1}\,dz$, with the contour enclosing $[-1,1]$. The resolvents are bounded on such a contour. The operator $D^{-1/2}B_SD^{1/2}=D^{-1/2}\,\Omega s^x\,D^{-1/2}$ is Hermitian, with Rayleigh quotient $\Omega\langle s^x\rangle/(\omega-\kappa\langle s^y\rangle)$. This lies in $[-1,1]$ because $(\langle s^x\rangle,\langle s^y\rangle)$ lies in the unit disk, so the spectrum of $B_S$ lies in $[-1,1]$ at every $S$. The similarity $D^{\pm1/2}$ is bounded uniformly in $S$, so $(z-B_S)^{-1}$ is bounded uniformly in $S$ for $z$ on any contour a fixed distance from $[-1,1]$, and by the argument above it has residual of order $1/S$. Integrating over the contour then gives the same for $f(B_S)A_S^n$. This proves the left eigenmodes of the main text satisfy
\begin{equation}
\Li_S^\dagger\rho^{L}_{n}=\ii n\Omega\,\rho^{L}_{n}+O(S^{-1}),\qquad \rho^{L}_{n}=f(B_S)A_S^{\,n}.\tag{\mnum{eq:general}}
\end{equation}
Because $\Li_S$ is not normal, a small residual does not by itself place an eigenvalue of $\Li_S$ near ${\rm i}n\Omega$. The dynamical statements in this work do not need this, and Fig.~\mnum{fig:one}(a,b) shows that the eigenvalues do converge, with the weight of $A_S$ outside the corresponding eigenspace vanishing as $1/S$.

\textit{Necessity.} If Eq.~\mref{eq:criterion} fails, $\lambda$ in Eq.~\eqref{eq:lambda} is either zero or has a nonzero real part, so there are no closed orbits: every trajectory that is not a fixed point flows into a fixed point as $t\to\infty$. Nothing can then oscillate indefinitely. Suppose $X_S$ has a continuous classical limit $y$ and oscillates at a frequency $\nu\neq0$, in the sense that $\|\Li_S^\dagger X_S-\ii\nu X_S\|\to0$. Because the evolution is a contraction, $e^{t\Li_S^\dagger}X_S=e^{\ii\nu t}X_S+o(1)$ at any fixed $t$. Mean-field dynamics is exact at fixed $t$ as $S\to\infty$~\cite{carollo21}, so the symbol of $e^{t\Li_S^\dagger}X_S$ is $y$ carried along the flow, and therefore $y(\bm n(t))=e^{\ii\nu t}y(\bm n(0))$ on every trajectory. At a fixed point $\bm n_*$ the trajectory never moves, so the oscillation condition reads $y(\bm n_*)=e^{\ii\nu t}y(\bm n_*)$ for all $t$, which for $\nu\neq0$ forces $y(\bm n_*)=0$. Along any other trajectory the same condition keeps $|y|$ constant, and since the trajectory flows into a fixed point, continuity forces this constant to be zero. So $y=0$ everywhere, and hence $\|X_S\|\to0$.

\textit{Completeness.} If Eq.~\mref{eq:criterion} holds, the same reasoning tells us which operators do oscillate. In the coordinates of the classical rotation $\xi$ is constant and the phase $\phi$ of $q=|q|e^{\ii\phi}$ advances uniformly, $\phi(t)=\phi(0)+\Omega t$. A function with $y(\bm n(t))=e^{\ii\nu t}y(\bm n(0))$ is then $y(\xi,\phi)=e^{\ii\nu\phi/\Omega}y(\xi,0)$, and it returns to itself when $\phi$ increases by $2\pi$ only if $\nu=k\Omega$ for an integer $k$. At the fixed points $|q|=0$ and $\phi$ is undefined, so for $y$ to be smooth there the factor $e^{\ii k\phi}$ must come with $|q|^{|k|}$, giving $y=f(\xi)q^k$ for $k\ge0$, or $f(\xi)\bar q^{|k|}$ for $k<0$, with $f$ smooth. So if $X_S$ has a smooth classical limit $y$ and oscillates at frequency $\nu$, as for necessity $\nu=k\Omega$ and $y=f(\xi)q^k$. Choosing a polynomial $p$ close to $f$ on $[-1,1]$, the operator $p(B_S)A_S^k$ has classical limit $p(\xi)q^k$, and since $\|\Op_S(a)\|\le\max|a|$ and $|q|\le1$, $\|X_S-p(B_S)A_S^k\|\le\max|f-p|+o(1)$. So $X_S$ is approximated arbitrarily well by the operators of Eq.~\mref{eq:general}, with $(A_S^\dagger)^{|k|}$ in place of $A_S^k$ for $k<0$.

\section{Stationary states}
\label{sec:ss}
This section derives the stationary state Eq.~\mref{eq:ss} and the right modes Eq.~\mref{eq:rightmodes} of a single spin, and then the dark state Eq.~\mref{eq:dark} and eigenmodes Eq.~\mref{eq:pairmodes} of the pair.

\textit{One spin.}
From here on we take the single jump of Eq.~\mref{eq:model}, written as $L=\sqrt{\kappa/S}\,c$ with $c=(\bm u+\ii\bm v)\cdot\bm S$.
Adding a constant $\gamma=S(a+\ii b)$, with $a$ and $b$ real, to the jump is equivalent to adding the term $\kappa(b\bm u-a\bm v)\cdot\bm S$ to the Hamiltonian. These Hamiltonians span the plane perpendicular to the dissipative axis $\bm u\times\bm v$. When Eq.~\mref{eq:criterion} holds the field lies in this plane, so a unique $\gamma$ absorbs the whole Hamiltonian and $\Li_S=\D[J]$ with $J=\sqrt{\kappa/S}\,(c+\gamma)$. For the jump of Eq.~\mref{eq:model} this is
\begin{equation}
J=\sqrt{\frac{\kappa}{S}}\,\bigl(\alpha S^-+\beta S^+ +\gamma\bigr),\tag{\mnum{eq:shiftedjump}}
\end{equation}
and for $\beta=0$ and $\bm h=\omega\hat{\bm x}$ it gives $\gamma=\ii\omega S/\kappa$. Any invertible $J$ then gives the stationary state~\cite{puri79,lourenco22}
\begin{equation}
\rho_{\rm ss}=\frac{(J^\dagger J)^{-1}}{\Tr\bigl((J^\dagger J)^{-1}\bigr)},\tag{\mnum{eq:ss}}
\end{equation}
since $J\rho_{\rm ss}J^\dagger$ and $\tfrac12\{J^\dagger J,\rho_{\rm ss}\}$ are both proportional to the identity and cancel in $\D[J]\rho_{\rm ss}$.

The main text also uses that $J$ is invertible with $\|J^{-1}\|=O(S^{-1/2})$. Every expectation value of $c/S$ lies in the filled ellipse $\{(\bm u+\ii\bm v)\cdot\bm n:|\bm n|\le1\}$, and so does the spectrum of $c/S$. The point $-\gamma/S=-(a+\ii b)$ lies a finite distance outside this ellipse, because $|\bm h|>\kappa|\bm u\times\bm v|$. Hence $\|J^{-1}\|=O(S^{-1/2})$. For $\beta=0$ the bound is explicit,
\begin{equation}
\|J^{-1}\|=\frac{\bigl\|(s^{-}+\ii\omega/\kappa)^{-1}\bigr\|}{\sqrt{\kappa S}}\le\frac{1}{(\omega/\kappa-1)\sqrt{\kappa S}}.
\end{equation}

\textit{Right modes.} The form $\Li_S=\D[J]$ also turns the left modes Eq.~\mref{eq:general} into the right modes Eq.~\mref{eq:rightmodes}. Using $\rho_{\rm ss}J^\dagger=J^{-1}/\Tr\bigl((J^\dagger J)^{-1}\bigr)$ and $\rho_{\rm ss}J^\dagger J\propto I$, a direct calculation gives, for any operator $X$, the exact identity
\begin{align}
\Li_S(X\rho_{\rm ss})=-(\Li_S^\dagger X)\rho_{\rm ss}+R_X\rho_{\rm ss}, \\R_X=\bigl[[J,X],J^\dagger\bigr]+[J,X]\,J^{-1}[J^\dagger,J].
\label{eq:rightid}
\end{align}
For $X=f(B_S)A_S^{\,n}$ each factor is controlled. By Eq.~\eqref{eq:commutator} and the resolvent argument used for the left modes, $[J,X]=\sqrt{\kappa/S}\,[c,X]$ is of order $S^{-1/2}$ and $[[J,X],J^\dagger]$ of order $S^{-1}$. The commutator $[J^\dagger,J]=\frac{\kappa}{S}[c^\dagger,c]$ is bounded, and $\|J^{-1}\|=O(S^{-1/2})$ by the bound above. So $\|R_X\|=O(S^{-1})$. Since $\|Y\rho_{\rm ss}\|_1\le\|Y\|$, Eq.~\eqref{eq:rightid} and Eq.~\mref{eq:general} give a trace-norm residual of order $1/S$,
\begin{equation}
\Li_S\rho^{R}_{n}=-\ii n\Omega\,\rho^{R}_{n}+O(S^{-1}),\qquad \rho^{R}_{n}=f(B_S)A_S^{\,n}\rho_{\rm ss}.\tag{\mnum{eq:rightmodes}}
\end{equation}

\textit{Two spins.} The pair of the main text is
\begin{equation}
\begin{aligned}
\Li_S^{(2)}\rho&=-\ii[H,\rho]+\frac{\Gamma}{S}\D[S_1^{-}+S_2^{-}]\rho,\\
H&=\omega(S_1^x+S_2^x)+\frac{g}{S}\bigl(S_1^xS_2^y-S_1^yS_2^x\bigr).
\end{aligned}\tag{\mnum{eq:pair}}
\end{equation}
The construction below applies equally to any jump $\alpha S^{-}+\beta S^{+}$ satisfying Eq.~\mref{eq:criterion}.

\textit{The dark state.} We work with two-spin states built from single-spin operators. The singlet
\begin{equation}
|s_0\rangle=\frac{1}{\sqrt{2S+1}}\sum_{m=-S}^{S}(-1)^{S-m}|m\rangle\otimes|{-m}\rangle,
\end{equation}
with $|m\rangle$ the eigenstates of $S^z$, is the unique state of the two spins with total spin zero. Write $\gamma=\ii\omega S/g$ for the single-spin constant with $\kappa$ replaced by $g$, so that $S^-+\gamma=S\,T_S$. Let $E=(S^-+\gamma)^{-1}=T_S^{-1}/S$ and $\Phi(O)=(OE\otimes I)|s_0\rangle$ for a single-spin operator $O$. The dark state of the main text,
\begin{equation}
|\Psi_S\rangle=\sqrt{\tfrac{2S+1}{Z}}\,\bigl(T_S^{-1}\otimes I\bigr)|s_0\rangle,\qquad T_S=s^{-}+\ii\frac{\omega}{g},\tag{\mnum{eq:dark}}
\end{equation}
is $|\Psi\rangle\propto\Phi(I)$~\cite{stannigel12,chu26}, and a general $\Phi(O)$ is that state with $O$ applied to the first spin. The aim is to show that the two-spin generator acting on $\Phi(O)$ is the single-spin generator acting on $O$, plus small corrections.

\textit{Moving operators between spins.} The jump of the pair is $C=S_1^-+S_2^-$. Since $\omega S^x=\frac{\ii g}{2S}(\bar\gamma S^{-}-\gamma S^{+})$ and $S_1^xS_2^y-S_1^yS_2^x=\frac{\ii}{2}(S_1^+S_2^{-}-S_2^+S_1^{-})$, the Hamiltonian is
\begin{equation}
H=\omega(S_1^x+S_2^x)+\frac{\ii g}{2S}\bigl(S_1^+ S_2^--S_2^+ S_1^-\bigr).
\end{equation}
Three facts reduce the two-spin generator to single-spin operations. First, the singlet has total spin zero, so $(I\otimes X)|s_0\rangle=-(X\otimes I)|s_0\rangle$ for any $X$ linear in the spin. This moves operators from the second spin onto the first. For linear $X$ and $Y$ it gives
\begin{equation}
\begin{aligned}
(X\otimes I+I\otimes X)\,\Phi(O)&=([X,OE]\otimes I)|s_0\rangle,\\
(X\otimes Y)\,\Phi(O)&=-(XOEY\otimes I)|s_0\rangle.
\end{aligned}
\label{eq:transfer}
\end{equation}
Second, $S^-$ commutes with $E$, while $[S^+,E]=FE$ with $F=E[S^-,S^+]$, which is bounded uniformly in $S$. Third, $[\omega S^x,E]=-\frac{\ii g\gamma}{2S}FE$.

The first line of Eq.~\eqref{eq:transfer} with $X=S^-$ and $X=S^+$, together with the second fact, gives
\begin{equation}
C\,\Phi(O)=\Phi([S^-,O]),\qquad C^\dagger\Phi(O)=\Phi([S^+,O]+OF).
\label{eq:Cid}
\end{equation}
Applying both, $C^\dagger C\,\Phi(O)=\Phi(\mathcal B(O))$ with $\mathcal B(O)=[S^+,[S^-,O]]+[S^-,O]F$. For the Hamiltonian we use both lines of Eq.~\eqref{eq:transfer}, then the three facts and $O(S^-+\gamma)E=O$ to move every factor of $E$ to the right. This gives
\begin{equation}
\begin{aligned}
-\ii H\Phi(O)=\Phi\Bigl(&-\ii[\omega S^x,O]+\tfrac{g}{2S}\bigl(S^-OS^+-S^+OS^-\bigr)\\
&-\tfrac{g}{2S}\bigl([S^-,O]F+O[S^-,S^+]\bigr)\Bigr).
\end{aligned}
\end{equation}
The argument equals $-\Li_1^\dagger O-\frac{g}{2S}\mathcal B(O)$, where $\Li_1$ is the single-spin Liouvillian with $\kappa\to g$. With the effective non-Hermitian Hamiltonian $K=-\ii H-\Gamma C^\dagger C/(2S)$ we therefore have
\begin{equation}
K\,\Phi(O)=\Phi\Bigl(-\Li_1^\dagger O-\tfrac{g+\Gamma}{2S}\mathcal B(O)\Bigr).
\label{eq:Kid}
\end{equation}
Setting $O=I$ gives $C|\Psi\rangle=0$ and $K|\Psi\rangle=0$, hence $H|\Psi\rangle=0$. The dark state is annihilated by both the jump and the Hamiltonian, so $|\Psi\rangle\langle\Psi|$ is stationary for every $\Gamma>0$. These identities are exact at every $S$.

\textit{Pair eigenmodes.} The same identities give the eigenmodes of the pair. Take $O=f(B_S)A_S^{\,n}$ and $O'=g(B_S)A_S^{\,m}$, left eigenmodes of the single spin with frequencies $n\Omega$ and $m\Omega$. Then $\mathcal B(O)$ and $\mathcal B(O')$ are bounded, and by Eq.~\eqref{eq:Kid} each side of $\Phi(O)\Phi(O')^\dagger$ evolves as its single-spin operator, so it is an eigenmode of the pair with frequency $-(n-m)\Omega$, up to a trace-norm residual of order $1/S$. The jump term $\Gamma C\rho C^\dagger/S$ contributes only at order $1/S$, because $C\Phi(O)=\Phi([S^-,O])$ is bounded. Since $O_1|\Psi\rangle\propto\Phi(O)$, these are the two-sided products of the main text,
\begin{equation}
\begin{aligned}
\Li_S^{(2)}\rho^{R}_{n,m}&=-\ii(n-m)\Omega\,\rho^{R}_{n,m}+O(S^{-1}),\\
\rho^{R}_{n,m}&=f(B_1)A_1^{\,n}\rho^{(2)}_{\rm ss}(A_1^\dagger)^{m}g(B_1)^\dagger.
\end{aligned}\tag{\mnum{eq:pairmodes}}
\end{equation}
The bath rate $\Gamma$ enters only through the terms of order $1/S$, so it sets how fast these modes decay but not their frequency.

\section{Relaxation of $B_S$}
\label{sec:diffusion}
As $S\to\infty$ every operator $f(B_S)\rho_{\rm ss}$ is stationary, but at finite $S$ only $\rho_{\rm ss}$ is. Here we derive, for $\beta=0$ and $\bm h=\omega\hat{\bm x}$, the distribution of $B_S$ in $\rho_{\rm ss}$ and how the rest of this continuum decays. Both results were found for the classical orbits by Drummond and Carmichael~\cite{drummond78,carmichael80}, and we give them here in operator form.

\textit{The stationary distribution of $B_S$.} We work in the coordinates $(\xi,\phi)$ of the classical rotation, where $q=\sqrt{1-\xi^2}\,e^{\ii\phi}$. Writing $x$ for the value of $\xi$ and $u={\rm Re}\,q=\sqrt{1-x^2}\cos\phi$, Eq.~\mref{eq:dressedsphere} inverts to
\begin{equation}
\bm n=\frac{\bigl(x\Omega,\;u\omega+\kappa,\;\sqrt{1-x^2}\,\Omega\sin\phi\bigr)}{\omega+\kappa u}.
\end{equation}
This holds whenever Eq.~\mref{eq:criterion} is satisfied, for any $\beta$, since $\xi$ and $q$ do not depend on $\beta$. In these coordinates the area element on the sphere is $d\bm n=\Omega^2\,dx\,d\phi/(\omega+\kappa u)^2$.

For $\beta=0$ the stationary state is $\rho_{\rm ss}\propto(J^\dagger J)^{-1}$ with $J\propto s^-+\ii\omega/\kappa$. Its coherent-state expectation value $\langle\bm n|\rho_{\rm ss}|\bm n\rangle$ is of order $1/d$, since $\Tr(\rho_{\rm ss})=1$, and as $S\to\infty$
\begin{equation}
d\,\langle\bm n|\rho_{\rm ss}|\bm n\rangle\to\frac{c_0}{|\hat J|^2}=\frac{c_0}{n_x^2+(\omega/\kappa-n_y)^2},
\end{equation}
with $c_0$ a normalisation constant. In the coordinates above, $n_x^2+(\omega/\kappa-n_y)^2=\Omega^2(\Omega^2+\kappa^2x^2)/[\kappa^2(\omega+\kappa u)^2]$. Its factor of $(\omega+\kappa u)^2$ cancels the one in the area element, and
\begin{equation}
\Tr\bigl(\rho_{\rm ss}f(B_S)\bigr)\to\int_{-1}^{1}p(x)f(x)\,dx,\quad p(x)\propto\frac{1}{\Omega^2+\kappa^2x^2},
\label{eq:pofx}
\end{equation}
as $S\to\infty$. So the stationary state is uniform in the phase $\phi$ and weights the values of $B_S$ with the density $p(x)$, as stated in the main text.

\textit{The slow dynamics.} An operator $f(B_S)$ changes only through the $O(1/S)$ terms of $\Li_S^\dagger$, so it evolves on times of order $S/\kappa$, over which the phase $\phi$ goes round many times. Only the average over $\phi$ then matters. The averaged generator depends on $\xi$ alone and is second order in $f$, so it is a drift and a diffusion,
\begin{equation}
\Li_S^\dagger f(B_S)\simeq(\mathcal Gf)(B_S),\qquad\mathcal Gf=\frac1S\bigl(mf'+\tfrac12wf''\bigr),
\end{equation}
with two functions $m(x)$ and $w(x)$ on $-1\le x\le1$. The part of $\Li_S^\dagger f(B_S)$ that depends on $\phi$ is of the same order but averages to zero, and it is removed by adding to $f(B_S)$ a correction of order $1/S$, found by dividing each Fourier component $e^{\ii n\phi}$ by $\ii n\Omega$. With this correction the relation above holds up to $O(S^{-2})$.

The diffusion comes from the jump. For a single jump $L$ the dissipator fails the product rule by exactly $[L,X]^\dagger[L,X]$, and for $X=f(B_S)$ and $L=\sqrt{\kappa S}\,s^-$ Eq.~\eqref{eq:commutator} turns this into $w|f'|^2/S$, with $w$ the orbit average of $\kappa|\{n^-,\xi\}|^2$, the Poisson bracket of Eq.~\eqref{eq:poisson}, and $n^-=n_x-\ii n_y$. In the coordinates above $|\{n^-,\xi\}|^2=(\Omega^2+\kappa^2\xi^2)(1-\xi^2)\sin^2\phi/\Omega^2$, so
\begin{equation}
w(x)=\frac{\kappa}{2\Omega^2}\bigl(\Omega^2+\kappa^2x^2\bigr)\bigl(1-x^2\bigr).
\end{equation}
The drift then follows from the stationary state. Since $\rho_{\rm ss}$ is exactly stationary, $\Tr\bigl(\rho_{\rm ss}\Li_S^\dagger f(B_S)\bigr)=0$ for every $f$, and by Eq.~\eqref{eq:pofx} this requires $\int p\,\mathcal Gf\,dx=0$. With no flux through the fixed points $x=\pm1$ this gives $mp=\frac12(wp)'$, and so
\begin{equation}
\mathcal Gf=\frac{(wpf')'}{2Sp}=\frac{\kappa}{S}\,\frac{\Omega^2+\kappa^2x^2}{4\Omega^2}\bigl[(1-x^2)f'\bigr]'.
\end{equation}
Its eigenvalues are $-\kappa\mu_k/S$, with eigenfunctions $f_k$ that solve the equation of the main text,
\begin{equation}
\frac{1}{4\Omega^2}\bigl(\Omega^2+\kappa^2x^2\bigr)\frac{d}{dx}\Bigl[(1-x^2)\,f_k'(x)\Bigr]=-\mu_k\,f_k(x).\tag{\mnum{eq:diffusion}}
\end{equation}
For $\kappa\ll\omega$ the factor $\Omega^2+\kappa^2x^2$ becomes constant, Eq.~\mref{eq:diffusion} becomes Legendre's equation, and $\mu_1\to1/2$. Because the corrected residual is $O(S^{-2})$ and the evolution is a contraction, $\Tr\bigl(\rho(t)f_k(B_S)\bigr)=e^{-\kappa\mu_kt/S}\,\Tr\bigl(\rho(0)f_k(B_S)\bigr)+O(S^{-1})$ for $t$ of order $S/\kappa$. The operator $\mathcal G$ is the period-averaged Fokker--Planck operator for the orbit label of Refs.~\cite{drummond78,carmichael80}, under the change of variable $x=(1-c)/(1+c)$ from their orbit label $c$. As for $A_S$, a small residual does not by itself place an eigenvalue of $\Li_S$ near $-\kappa\mu_k/S$, but numerically, for $\omega=2\kappa$ and $S$ up to $160$, the slowest non-oscillating eigenvalue converges to $-\kappa\mu_1/S$ and its left eigenmode differs from $f_1(B_S)$ by $O(1/S)$.

\textit{The correlation of $B_S$.} The generator $\mathcal G$ is self-adjoint with respect to the weight $p$, so its eigenfunctions can be chosen orthonormal, $\int p\,f_jf_k\,dx=\delta_{jk}$. Expanding the symbol $x$ of $B_S$ in them gives, at fixed $\tau=\kappa t/S$,
\begin{equation}
\lim_{S\to\infty}{\rm Tr}\bigl(\rho_{\rm ss}B_S^\dagger B_S(t)\bigr)=\sum_{k\ge1}|b_k|^2e^{-\mu_k\tau},
\end{equation}
with $b_k=\int_{-1}^{1}p\,x\,f_k\,dx$, where the $k=0$ term vanishes because $p$ is even. The function $x$, the symbol of $B_S$, is almost exactly the first of them, $f_1$, with over $99\%$ of its weight on $f_1$ even at $\omega=1.2\kappa$. The sum is therefore a single exponential to excellent accuracy,
\begin{equation}
\lim_{S\to\infty}{\rm Tr}\bigl(\rho_{\rm ss}B_S^\dagger B_S(t)\bigr)\simeq\Bigl(\frac{\Omega}{\kappa\theta}-\frac{\Omega^2}{\kappa^2}\Bigr)e^{-\kappa\mu_1t/S},
\end{equation}
with $B_S(t)=e^{t\Li_S^\dagger}B_S$, $\theta=\arctan(\kappa/\Omega)$ and the limit taken at fixed $\kappa t/S$. The amplitude is $\langle\xi^2\rangle=\int p\,x^2\,dx$, which is one minus the amplitude of the correlation of $A_S$ in the main text,
\begin{equation}
\lim_{S\to\infty}{\rm Tr}\bigl(\rho_{\rm ss}A_S^\dagger A_S(t)\bigr)=\Bigl(\frac{\omega^2}{\kappa^2}-\frac{\Omega}{\kappa\theta}\Bigr)e^{\ii\Omega t},\tag{\mnum{eq:correlation}}
\end{equation}
as it must be since $A_S^\dagger A_S=1-B_S^2+O(S^{-1})$. The full sum gives the dashed line in Fig.~\mnum{fig:one}(c), and the exact finite-$S$ correlation approaches it with deviations that fall as $1/S$.

\section{Entanglement}
\label{sec:trace}
The logarithmic negativity~\cite{vidal02,plenio05} of a state $\rho$ of the two spins is $\EN(\rho)=\log\|\rho^{T_2}\|_1$, where $\rho^{T_2}$ is the partial transpose on the second spin and $\|\cdot\|_1$ is the trace norm. For a pure state with Schmidt coefficients $\sqrt{p_i}$ it reduces to $\EN=2\log\sum_i\sqrt{p_i}$. For the state $(O\otimes I)|\Psi_S\rangle$ these are proportional to the singular values of the single-spin matrix $M=OT_S^{-1}$, so
\begin{equation}
\EN=2\log\Tr\bigl((M^\dagger M)^{1/2}\bigr)-\log\Tr\bigl(M^\dagger M\bigr).
\end{equation}
This requires traces of functions of $M^\dagger M$ at large $S$. We can do this by observing that the trace over the $d$-dimensional space becomes $d$ times an average over the sphere, with each operator replaced by its symbol. The following lemma makes this precise.

\begin{lemma}
\label{lem:trace}
Let $X_S=P_SR_S^{-1}$ with $P_S$ and $R_S$ fixed polynomials in $\bm s$, and suppose that the symbol $\hat R_S$ is nowhere zero on the sphere and that $R_S^{-1}$, $X_S$ and $X_S^{-1}$ are bounded uniformly in $S$. Then for any $f$ holomorphic near the positive interval that contains the spectrum of $X_S^\dagger X_S$ and the range of $|\hat X_S|^2$,
\begin{equation}
\tfrac1d\Tr\bigl(f(X_S^\dagger X_S)\bigr)=\overline{f(|\hat X_S|^2)}+O(S^{-1}).
\label{eq:trace}
\end{equation}
\end{lemma}
\begin{proof}
Since $\hat R_S$ is nowhere zero, $1/\hat R_S$ is smooth and Eq.~\eqref{eq:product} gives $R_S\Op_S(1/\hat R_S)=I+O(S^{-1})$. Multiplying by the bounded $R_S^{-1}$ gives $R_S^{-1}=\Op_S(1/\hat R_S)+O(S^{-1})$. Hence $X_S=\Op_S(\hat X_S)+O(S^{-1})$ and $X_S^\dagger X_S=\Op_S(|\hat X_S|^2)+O(S^{-1})$. The same argument gives $(z-X_S^\dagger X_S)^{-1}=\Op_S\bigl((z-|\hat X_S|^2)^{-1}\bigr)+O(S^{-1})$ for every $z$ on a contour around the interval, since $X_S^\dagger X_S$ is Hermitian and so its resolvent is bounded by the inverse distance from $z$ to the interval. Cauchy's integral formula writes $f(X_S^\dagger X_S)$ as an integral of these resolvents, so $f(X_S^\dagger X_S)=\Op_S\bigl(f(|\hat X_S|^2)\bigr)+O(S^{-1})$. Taking the trace, with $\Tr\bigl(\Op_S(a)\bigr)=d\,\overline a$, proves the lemma.
\end{proof}

If $O$ has a smooth symbol that is bounded away from zero, $M$ satisfies Lemma~\ref{lem:trace} with $\hat M=\hat O/\hat T_S$, where $\hat T_S=n_x-\ii n_y+\ii\omega/g$ is the symbol of $T_S$. Applying Lemma~\ref{lem:trace} with $f(x)=x^{1/2}$ and $f(x)=x$,
\begin{equation}
\EN=\log d+2\log\overline{|\hat M|}-\log\overline{|\hat M|^2}+O(S^{-1}),
\label{eq:ENb}
\end{equation}
For the stationary state $O=I$ and $|\hat M|^{-2}=n_x^2+(\omega/g-n_y)^2$. Both averages are elementary in the coordinates $(x,\phi)$ introduced for the stationary distribution of $B_S$, with $g$ in place of $\kappa$, and give $\overline{|\hat M|}=\theta$ and $\overline{|\hat M|^2}=g\theta/\Omega$ with $\theta=\arctan(g/\Omega)$. This is the result of the main text,
\begin{equation}
\EN(\rho^{(2)}_{\rm ss})=\log(2S+1)+\log\Bigl(\frac{\Omega\,\theta}{g}\Bigr)+O(S^{-1}).\tag{\mnum{eq:ENss}}
\end{equation}

\section{Quench dynamics}
\label{sec:quench}
The quench reduces to a single-spin problem: the initial state is a single-spin operator $O$ applied to the new dark state, the pair carries $O$ along the single-spin orbits, and Eq.~\eqref{eq:ENb} turns the resulting Schmidt matrix into the negativity. Quenching from $g'$ to $g$ starts the pair in the dark state for $g'$. Both dark states come from the same singlet, $|\Psi'_S\rangle\propto(T'^{-1}_S\otimes I)|s_0\rangle$ and $|\Psi_S\rangle\propto(T^{-1}_S\otimes I)|s_0\rangle$, so the initial state is $(O\otimes I)|\Psi_S\rangle$ with $O=T'^{-1}_ST_S$. Its symbol is $\hat O=\hat T_S/\hat T'_S$, which is bounded away from zero, $|\hat O|\ge(\omega/g-1)/(\omega/g'+1)$, so Lemma~\ref{lem:trace} applies to $M=OT_S^{-1}$.

Under the new coupling the operator $O$ evolves as a single-spin operator, by Eq.~\eqref{eq:Kid}, and its symbol is carried along the classical orbits of the coupling $g$. In the coordinates of Eq.~\mref{eq:dressedsphere} this is the shift $\phi\to\phi-\Omega t$. At time $t$ the symbol of $M$ is therefore
\begin{equation}
\hat M_t(\bm n)=\frac{\hat T_S(\bm n_t)}{\hat T'_S(\bm n_t)\,\hat T_S(\bm n)},
\end{equation}
where $\bm n_t$ is $\bm n$ carried back along its orbit by $\Omega t$. The two factors of $\hat T_S$ do not cancel, because they are evaluated at different points. To evaluate the averages in Eq.~\eqref{eq:ENb} we use the coordinates $(x,\phi)$ introduced for the stationary distribution of $B_S$, with $g$ in place of $\kappa$, where the sphere average becomes
\begin{equation}
\overline{(\,\cdot\,)}=\frac{\Omega^2}{4\pi}\int_{-1}^{1}\!dx\int_0^{2\pi}\!\frac{d\phi}{D(x,\phi)^2}\,(\,\cdot\,),
\end{equation}
with $r=\sqrt{1-x^2}$ and
\begin{equation}
\begin{aligned}
D(x,\phi)&=\omega+gr\cos\phi,\\
Q(x,\phi)&=\bigl[\omega^2-gg'+\omega(g-g')r\cos\phi\bigr]^2+g'^2\Omega^2x^2.
\end{aligned}
\end{equation}
In these coordinates
\begin{equation}
|\hat M_t|=\frac{g'D(x,\phi)}{\sqrt{Q(x,\phi-\Omega t)}},
\end{equation}
so $\overline{|\hat M_t|}$ is
\begin{equation}
\mathcal N(t)=\frac{g'\Omega^2}{4\pi}\int_{-1}^{1}\!dx\int_0^{2\pi}\!\frac{d\phi}{D(x,\phi)\sqrt{Q(x,\phi-\Omega t)}}.\label{eq:N}
\end{equation}
Shifting $\phi$ shows that $\overline{|\hat M_t|^2}$ does not depend on time, so it keeps its value at $t=0$, where $M=T'^{-1}_S$ and the stationary result gives $g'\theta'/\Omega'$, with $\Omega'=\sqrt{\omega^2-g'^2}$ and $\theta'=\arctan(g'/\Omega')$. Eq.~\eqref{eq:ENb} then becomes
\begin{equation}
\EN(\rho(t))=\log(2S+1)+\log\Bigl(\frac{\Omega'\,\mathcal N(t)^2}{g'\theta'}\Bigr)+O(S^{-1}).\label{eq:ENtgen}
\end{equation}

For a quench from the singlet, $g'\to0$, we have $|\hat M_t|=D(x,\phi)/D(x,\phi-\Omega t)$ and $\overline{|\hat M_t|^2}=1$. Combining the two denominators of $\overline{|\hat M_t|}$ with a Feynman parameter makes the remaining integrals elementary, and
\begin{equation}
\mathcal N(t)=\frac{\sinh^{-1}u_t}{u_t\sqrt{1+u_t^2}},\qquad u_t=\frac{g}{\Omega}\Bigl|\sin\frac{\Omega t}{2}\Bigr|.
\end{equation}
Eq.~\eqref{eq:ENb} then gives the result of the main text,
\begin{equation}
\begin{aligned}
\EN(\rho(t))&=\log(2S+1)\\
&\quad+2\log\left(\frac{\sinh^{-1}(u_t)}{u_t\sqrt{1+u_t^2}}\right)+O(S^{-1}),
\end{aligned}\tag{\mnum{eq:ENt}}
\end{equation}
with $u_t$ as above.

\textit{Validity.} These results use the fact that the pure state $\Phi(O_t)$ solves, on any fixed time window, the full master equation up to $O(1/S)$ in trace norm. The jump term is small because $C\,\Phi(O_t)$ is bounded. Partial transposition can enlarge this error by at most a factor $d$, but $\|\Phi(O_t)^{T_2}\|_1$ is itself of order $d$, so Eqs.~\eqref{eq:ENtgen} and \mref{eq:ENt} hold up to $O(S^{-1})$ for fixed $t$ and $g$ away from $\omega$, though not for times of order $S$.

\section{AI Usage}
Frontier large language models (ChatGPT's Astra and Anthropic's Claude Opus 5.5) were used to derive mathematical material in this Supplemental Material. This was checked, simplified and re-written in places by the author, who takes full responsibility for its content. The starting point was prompting by the author to identify the key quantum operators that could be applied to the known stationary state to generate imaginary eigenmodes as $S \rightarrow \infty$.

\bibliography{main}